\documentclass[%
reprint,
amsmath,amssymb,
aps,
pre,
floatfix,
]{revtex4-2}

\usepackage[utf8]{inputenc} 
\usepackage[T1]{fontenc}    
\usepackage[colorlinks=true,citecolor=blue]{hyperref}       
\usepackage{url}            
\usepackage{booktabs}       
\usepackage{amsfonts}       
\usepackage{nicefrac}       
\usepackage{microtype}      
\usepackage{graphicx}
\usepackage{amsmath}
\usepackage{amssymb}
\usepackage{comment}
\usepackage[dvipsnames]{xcolor}

\DeclareMathOperator{\sign}{sign}

\newcommand{\bxi}{\boldsymbol{\xi}}
\newcommand{\bmu}{\boldsymbol{\mu}}
\newcommand{\bss}{\boldsymbol{s}}
\newcommand{\bbf}{\boldsymbol{f}}

\newcommand{\bc}{\boldsymbol{c}}
\newcommand{\cH}{\mathcal{H}}

\colorlet{ochre}{blue!30!yellow!70!}

\newcommand{\vect}[1]{\mathbf{#1}}

\DeclareMathOperator\erf{erf}

\begin{document}

\title[]{Random Features Hopfield Networks generalize retrieval to previously unseen examples}

\author{Silvio Kalaj$^1$,
Clarissa Lauditi$^2$,
Gabriele Perugini$^{3,4}$,
Carlo Lucibello$^{3,4}$,
Enrico M. Malatesta$^{3,4}$,
Matteo Negri$^{1,5}$}
\email[Corresponding author; ]{matteo.negri@uniroma1.it}

\affiliation{$^1$ University of Rome ‘La Sapienza’, Department of Physics, Piazzale Aldo Moro 5, 00185 Roma, Italy}
\affiliation{$^2$ Department of Applied Math, John A. Paulson School of Engineering and Applied Sciences, Harvard University, 02138 Cambridge, MA, USA}
\affiliation{$^3$ Department of Computing Sciences, Bocconi University, 20136 Milano, Italy}
\affiliation{$^4$ Institute for Data Science and Analytics, Bocconi University, 20136 Milano, Italy}
\affiliation{$^5$ CNR-NANOTEC, Institute of Nanotechnology, Rome Unit, Piazzale Aldo Moro, 00185 Roma, Italy}

\date{\today}

\begin{abstract}
  It has been recently shown that a learning transition happens when a Hopfield Network stores examples generated as superpositions of random features, where new attractors corresponding to such features appear in the model. In this work we reveal that the network also develops attractors corresponding to previously unseen examples generated with the same set of features. We explain this surprising behaviour in terms of spurious states of the learned features: we argue that, increasing the number of stored examples beyond the learning transition, the model also learns to mix the features to represent both stored and previously unseen examples.
  We support this claim with the computation of the phase diagram of the model.
\end{abstract}

\maketitle

The Hopfield Model \cite{hopfield1982neural} is a paradigmatic model of associative memory with relevance in physics, biology, and computer science. Starting from corrupted signals, stored binary memories are retrieved as fixed points of a dynamical system which is also an energy minimization process.  
Recently, generalizations of the Hopfield Model have gained attention thanks to the addition of several desirable properties that nonetheless preserve the energy minimization and the associative mapping paradigms. In particular, Modern Hopfield Networks overcome the linear (in the system size) capacity limit of the Hopfield Model and are able to store a polynomial~\cite{gardner1987multiconnected, krotov2016dense, agliari2023dense} or even exponential~\cite{demircigil2017model,ramsauer2020hopfield, lucibello2023exponential} number of memories. Continuous variables and differentiable update rules allow to plug in trainable components in machine learning applications \cite{ramsauer2020hopfield, krotov2023new}. A Lagrangian formalism can be used to describe a large family of such models \cite{krotov2021large}, also accommodating popular deep learning components such as the attention mechanism \cite{vaswani2017attention} and layer normalization \cite{ba2016layer}.

In connection with recent advances in theoretical machine learning \cite{mei2022generalization,goldt2020modeling,gerace2020generalisation}, it is shown in Ref. \cite{negri2023hidden} that even the standard Hopfield Model, which has no trainable parameters, when given memories generated by a latent manifold, $\bxi_\nu = \sigma(F\bc_\nu)$, is able to "learn" essential features of the data generating process: when provided with enough samples, the (unobserved) columns of $F$ become attractors of the dynamics. 
In this work, we further characterize the Random Features Hopfield Model of Ref. \cite{negri2023hidden}. In particular, we show that the model is also able to "generalize", that is to store unobserved samples of the data manifold. Spurious minima (mixtures in particular), normally detrimental for the Hopfield Model, become beneficial in this context.

\paragraph*{Random Features Hopfield Model.} 
Given $N$ binary neurons $s_i=\pm1$, and $P$ binary memories $\{\bxi_\nu\}_{\nu=1}^P$ that we want to store, the Hopfield Model defines the sequential update rule 
\begin{equation}
    s_i^{(t+1)} = \sign \left( \sum_{j\neq i}^N J_{ij} s_j^{(t)}\right),\quad  J_{ij}=\frac{1}{N}\sum_{\nu=1}^{P}\xi_{\nu i}\xi_{\nu j}.
    \label{eq:update_rule}
\end{equation}

It can be shown \cite{hopfield1982neural} that
as long as the memories are few (at most $O(N)$) and far apart enough, they approximately correspond to fixed points of the dynamics and can be retrieved from a perturbed configuration. The model admits an energy function $\cH(\bss) =-\frac{1}{2}\sum_{i\neq j}J_{ij}s_{i}s_{j}$. 

The standard Hopfield Model setting considers random uncorrelated memories with uniform components. Using statistical physics techniques \cite{amit1987statistical,amit1987information}, it has been shown that the model is able to store up to $P \approx 0.138 N$ memories for large $N$.

The Random Features Hopfield Model \cite{negri2023hidden} keeps the same update rule and couplings of eq.~\ref{eq:update_rule}, but considers a data structure given by a random projection of a $D$-dimensional latent space~\cite{rahimi2007random,goldt2020modeling,gerace2020generalisation,Baldassi2022Learning}:
\begin{equation}
    \bxi_{\nu}=\sign\left(F \bc_{\nu}\right).
    \label{eq:correlated_examples}
\end{equation}
The matrix $F\in\{-1,+1\}^{N\times D}$ has i.i.d. uniform components. We call \emph{features} its columns $\bbf_k$. The latents vectors $\bc_\nu \in\mathbb{R}^D$ are called \emph{coefficients} instead. We take each $\bc_\nu$ to have exactly $L$ non-zero entries, in random locations and uniformly sampled in $\pm1$.
We will discuss two cases: the sparse case, $L=O(1)$ as $D\to\infty$, and the fully dense case, $L=D$.

In the dense case Ref. \cite{negri2023hidden} shows that when a large number of examples is given to the Random Features Hopfield Model, beyond its storage capacity, it enters a phase where features instead become attractors. The model enters the learning phase if $\alpha=P/N$ is larger than a critical value that depends on $\alpha_D=D/N$. We refer to this as the \emph{learning transition}.


Given that the examples are correlated, it is useful to call \emph{training example}, $\bxi^\mathrm{train}$, any of the $P$ examples that are used in the Hebb rule defining the couplings matrix $J$ in \eqref{eq:update_rule}. We also call \emph{test example}, $\bxi^\mathrm{test}$, any linear combination (followed by the $\sign$ activation) of the features that is not used in $J$.
Note that in principle we could have different values of $L$ in training and test examples. To keep the numerical analysis simple, we only consider $L_\mathrm{train}= L_\mathrm{test}=L$.
We can now say that an Hopfield Model \emph{generalizes} if test examples are fixed points of the update rule~\eqref{eq:update_rule}. In the same spirit of the learning phase in \cite{negri2023hidden}, we study whether the network enters a \emph{generalization phase} in some region of $\alpha$, $\alpha_D$ and $L$ space.

\begin{figure}
    \centering
    \includegraphics[trim=0.3cm 0.3cm 0.0cm 0.0cm,clip,width=0.48\textwidth]{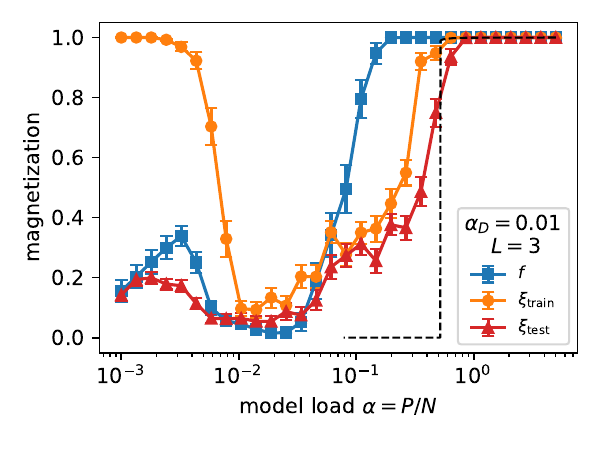}
    \caption{\textbf{Training and test examples become fixed points after the features have been learned.} Magnetization as a function of $\alpha$, for fixed $\alpha_D$. The {\textcolor{MidnightBlue}{\bf blue}} line is the magnetization $\mu$ of hidden features, which grows to $1$ if $\alpha$ is high enough (\emph{learning phase}). The {\textcolor{Orange}{\bf orange}} line is the magnetization $m^\mathrm{train}$ of the training examples, which is $m^\mathrm{train}\simeq1$ for low $\alpha$ and  drops when $\alpha$ increases, as expected from an associative memory (\emph{storage phase}). Surprisingly, $m^\mathrm{train}$ grows to $1$ again for high values of $\alpha$. Near this transition, also test examples have $m^\mathrm{test}=1$, as shown by the {\textcolor{red}{\bf red}} line (\emph{generalization phase}). $N=32000$; averages of $40$ samples. The dashed line shows the analytical prediction for the magnetization of mixtures of $3$ features.
    }
    \label{fig:main}
 \end{figure}

\paragraph*{Generalization transition.} 
The numerical results of this work are all measures of \emph{magnetizations}, obtained in the following way. First, we initialize the model to the configuration $\bss^{(0)}$ whose stability we want to check (we consider $\bss^{(0)}=\bxi^\mathrm{train}$, $\bss^{(0)}=\bxi^\mathrm{test}$, and $\bss^{(0)}=\bbf$, with $\bbf$ any of the columns of $F$). Then, we run the update rule~\eqref{eq:update_rule} until we reach a fixed point $\Tilde{\bss}$. Finally, we compute the magnetization as the normalized scalar product between the fixed point and the initial condition: $m^\mathrm{train}=\frac{1}{N}\sum_{i=1}^N \Tilde{s}_i \xi_i^\mathrm{train}$, $m^\mathrm{test}=\frac{1}{N}\sum_{i=1}^N \Tilde{s}_i \xi_i^\mathrm{test}$, and $\mu=\frac{1}{N}\sum_{i=1}^N \Tilde{s}_i f_i$ respectively. If, for given $\alpha$ and $\alpha_D$, we find that the magnetization is close to $1$, we say that there is a fixed point of eq.~\eqref{eq:update_rule} that corresponds  to $\bxi^\mathrm{train}$, $\bxi^\mathrm{test}$ or $\mathbf{f}$ respectively.

\begin{figure*}
    \centering
    \includegraphics[trim={0.4cm 0.2cm 0.0cm 0.0cm},width=0.49\textwidth,clip]{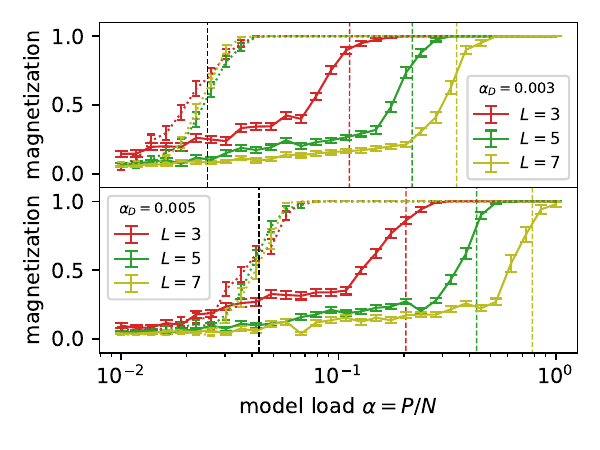}
    \includegraphics[
    trim={0.4cm 0.2cm 0.0cm 0.0cm},
    width=0.49\textwidth,clip]{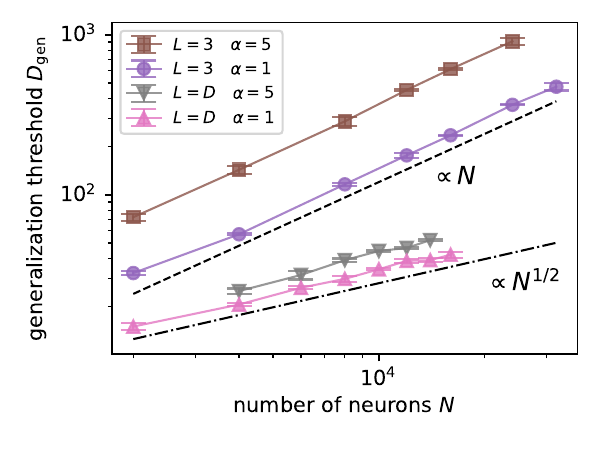}
    \caption{\textbf{ Combinations of more features require more training examples.} a) Feature magnetization $\mu$ ($dotted$) and test examples magnetization $m^\mathrm{test}$ ($solid$) as a function of $\alpha$ for different $\alpha_D$ (subplots).  Different colors represent increasing features per example $L$.
    Dashed vertical lines are the analytical predictions: in black we show the learning transition; colors correspond to the transitions for mixtures of $L=3,5,7$ features. ($N=32000$; averages of $40$ samples.) b) Scaling with $N$ of the maximum number of features $D_\mathrm{gen}$ for which we observe a generalization transition. 
    Specifically, we plot the maximum $D(N)$ at which $10$ samples have $m^\mathrm{test} > 0.9$, and we average over $4$ to $10$ groups of $10$ samples, depending on $N$.
    }

    \label{fig:manyL}
\end{figure*}

The first result that we present is that the model shows a learning transition even when it is trained with a sparse combination of features (see Fig.~\ref{fig:main}, blue line), extending the result of \cite{negri2023hidden}. The position of the learning transition depends weakly (if at all) on the number $L$ of features per example (see dotted lines in Fig.~\ref{fig:manyL}a).
 
Surprisingly, after the features have been learned, the model enters a phase in which the training examples are stable again (see Fig.~\ref{fig:main}, orange line). We note that it is physically implausible that a phase disappears for low $\alpha$ and reappears when $\alpha$ is large. To understand what is happening, we check the magnetization of test examples: we find that they become fixed points together with the training examples. We call this the \emph{generalization phase}, as it resembles the behavior of inference models that perform well on previously unseen examples. The fact that both train and test examples are fixed points for high $\alpha$, while only training examples are fixed points for low $\alpha$, indicates that the model must be using different mechanisms to achieve these results. 

In Fig.~\ref{fig:manyL}a we show 
generalization transitions (solid lines) for different values of $L$ and $\alpha_D$. We see that increasing either $L$ or $\alpha_D$ has the effect of moving the generalization transition to higher values of $\alpha$.
Additionally, in Fig.~\ref{fig:manyL}b we study the maximum number of features $D_\mathrm{gen}$ compatible with a generalization transition for a given value of $\alpha$. We see that is $D_\mathrm{gen}=O(N)$ in the sparse case $L=O(1)$  and $D_\mathrm{gen}=O(\sqrt{N})$ in the dense case $L=D$. This scaling is compatible with the signal-to-noise analysis discussed below.

\label{sec:theory}

\paragraph*{Theoretical results.}
We now provide a theoretical framework to explain how the Random Features Hopfield Model stores examples that it never saw before. 
First, we provide a simple signal-to-noise argument to get an intuition on why denser combination require more training examples, as shown in Fig.~\ref{fig:manyL}. This argument shows that we can interpret the generalization transition for a given $L$ as the mixtures of $L$ features becoming a fixed point of \ref{eq:update_rule}.
Then, we test this interpretation by resorting to the statistical physics formulation of the Random Features Hopfield Model \cite{negri2023hidden} and computing the spinodal lines of mixtures of features. Note that, to do this, we consider dense train examples $L_\mathrm{train}=D$, since treating sparse training examples would be much harder and is outside of the scope of this work. This simplification is justified by the surprising numerical result that the learning transition depends weakly on $L$ (see Fig.~\ref{fig:manyL}a). These assumpions prove to be consistent, as we see that the spinodal lines predict accurately the generalization transitions (see Fig.~\ref{fig:main} and Fig.~\ref{fig:manyL}a). 

\paragraph*{Signal-to-noise analysis.} 
\label{sec:signal-to-noise}
Let us define a mixture of features as 
$
    \chi_{\nu i}=\sum_{k''}c_{\nu k''} f_{k'' i},
$
where at this level the coefficients can be either dense or sparse (for simplicity we set $\sigma(x)=x$ just for this analysis). The local field on a mixture can be written (see appendix \ref{app:signal-to-noise} in supplemental material (SM)) as a signal term, proportional to the mixture itself, plus "crosstalk" terms that include the noise coming from the other features:
\begin{multline}
\label{eq::signaltonoise-mixture}
\sum_{j=1}^N J_{ij}\chi_{\nu j}  = \frac{P}{D}\left[\chi_{\nu i}+O\left(\sqrt{\frac{LD}{N}}\right)\right. \\
 \left.+O\left(\sqrt{\frac{L^{3}}{DP}}\right)+O\left(\sqrt{\frac{L^{3}}{NP}}\right)\right].
\end{multline}
Now we can see that, if $L=O(1)$, the second and third noise terms go to zero when $N, D\to\infty$; then, to keep the first noise term finite, we must set $D=O(N)$. This means that the retrieval of (sparse) mixtures happens at $\alpha>0$. On the contrary, it is impossible to retrieve (dense) mixtures at finite $\alpha$ when $L=D=O(N)$, since the signal would be overwhelmed by noise. It becomes possible instead when $L=D=O(N^{1/2})$, meaning $\alpha=0$: in this regime, the first noise term is finite again (as well as the second and third ones).
This scaling is confirmed by numerical results (see Fig.~\ref{fig:manyL}b).

\paragraph*{Replica analysis.} 
We extend the calculation reported in \cite{negri2023hidden} for the single-feature retrieval to the case of retrieval of mixtures of features. Here we sketch the main steps, describing the full derivation in appendix \ref{app:replica_calculation} (SM).
We are interested in predicting the location of the phase transition characterized by a discontinuity of the feature magnetizations $\bmu=\{\mu^1, ... ,\mu^D\}$. Given that we have one order parameter for each feature, we need to make an ansatz on the structure of the magnetizations to make the problem treatable. The single-feature retrieval consists in looking in a solution of the form $\bmu=\{\mu,0, ... ,0\}$, where one magnetization remains finite and the others vanish in the thermodynamic limit. Here, instead we consider $L$ finite magnetizations, namely $\bmu=\{\mu_1,...,\mu_L,0, ... ,0\}$.
To identify phase transitions we write the average free energy
\begin{equation}
    \phi = \lim_{N\to\infty} -\frac{1}{\beta N} \langle \ln Z \rangle_{c,f},
    \label{eq:phi}
\end{equation}
where $Z=\sum_{\bss} \exp(-\beta \mathcal{H}(\bss))$ is the partition function and $\langle ... \rangle_{c,f}$ is the average over the coefficients $c$ and the features $f$. To compute $\langle \ln Z \rangle_{c,f}$ we use the replica method \cite{mezard1987spin}, namely we represent the logarithm as  $\langle \ln Z \rangle = \lim_{n\to0} (\langle Z^n \rangle-1)/n$.
After inserting delta functions for the definition of the example magnetizations $m^a_\nu=\frac{1}{N}\sum_i \xi_{\nu i} s_i^a$, the replicated partition function reads
\begin{multline}
    \langle  Z^{n}\rangle = e^{-\frac{\beta}{2}Pn}\sum_{\left\{ s_{i}^{a}\right\} }\int\prod_{\nu a}\frac{dm_{\nu}^{a}}{\sqrt{2\pi}}
    e^{ \frac{\beta}{2}\sum_{\nu=1}^{P}\sum_{a=1}^{n}\left(m_{\nu}^{a}\right)^{2}} \\ 
    \times \left\langle \prod_{\nu a}\delta\left(m_{\nu}^{a}-\frac{1}{N}\sum_{i=1}^{N}\sigma\left(\frac{1}{\sqrt{D}}\sum_{k=1}^{D}c_{\nu k}f_{ki}\right)s_{i}^{a}\right)\right\rangle _{c,f}. \label{eq:rfhmpartitionfunctioninit}  
\end{multline}
Following \cite{negri2023hidden}, to proceed we need to isolate the finite terms inside the function $\sigma$. Without loss of generality we can assume that the non-vanishing magnetizations correspond to the first $L$ values of the index $k$, so that we integrate away the coefficients $c_{L+1},...,c_D$. Compared with the single-retrieval calculation, the resulting distribution of $m^a_\nu$ is also a Gaussian $\mathcal{N}(m_{\nu};\bar{m},Q)$ with the same covariance $Q$, but a different mean 
$
    \bar{m} = \frac{\kappa_1}{\sqrt{\alpha_D}}\sum_{k=1}^L c_{\nu k}\mu_k^a,
$
where $\kappa_{1}  =\int \frac{dz}{\sqrt{2\pi}} e^{-z^2/2}\,z\,\sigma(z)$.
\begin{figure}
    \centering
    \includegraphics[trim=0.8cm 0.9cm 1.2cm 2.0cm,clip,width=0.99\columnwidth]{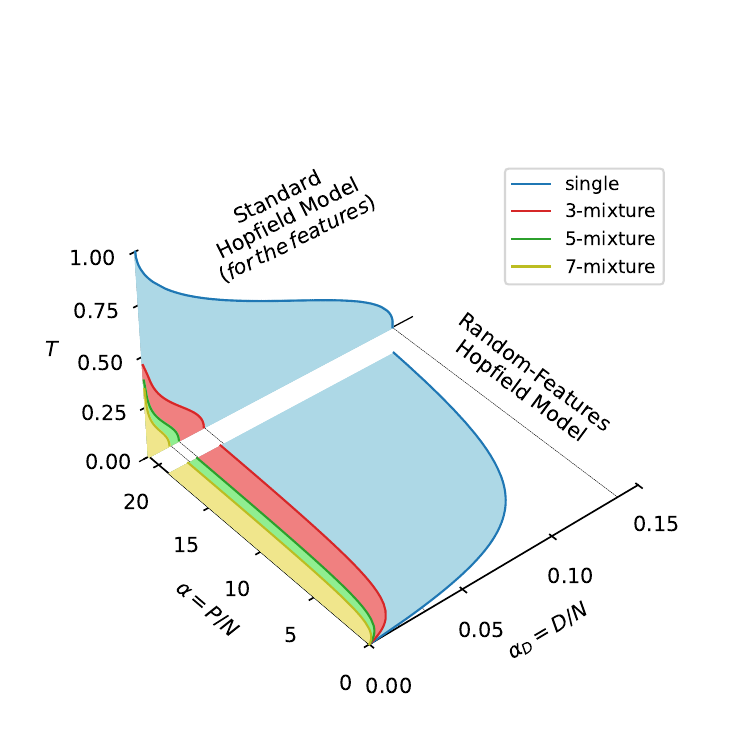}
    \caption{\textbf{Comparison between the phase diagrams} of a standard Hopfield Model from \cite{amit1987statistical} (\textit{top}, temperature $T$ vs $\alpha$) and the phase diagram of the dense Random Features Hopfield Model (\textit{bottom}, $\alpha$ vs $\alpha_D$). In both panels, the {\textcolor{MidnightBlue}{\bf blue}} line is the retrieval line, below which the features can be stored and retrieved. The {\textcolor{red}{\bf red}}, {\textcolor{ForestGreen}{\bf green}} and {\textcolor{ochre}{\bf yellow}} lines are the retrieval lines of mixtures, respectively of 3, 5 and 7 examples.
    When $\alpha\to\infty$, the curves of the Random Features Hopfield Model connect to the critical points at $T=0$ in the Hopfield Model: high order mixtures are stable at lower $\alpha_D$. We also show how these critical points evolve at finite temperature.
    }
    \label{fig:phase_diagram}
\end{figure}
See appendix \ref{app:gaussian_equivalence} (SM) for the explicit derivation, which can also be seen as a specialization of the Gaussian Equivalence Theorem \cite{mei2022generalization, gerace2020generalisation, goldt2022gaussian, goldt2020modeling, hu2022universality,Baldassi2022Learning}.

We end up with the following expression for the replicated partition function:
\begin{multline}
	\left\langle Z^{n}\right\rangle =e^{-\frac{\beta}{2}Pn}\sum_{\left\{ s_{i}^{a}\right\} }\int\prod_{\nu a}\frac{dm_{\nu}^{a}}{\sqrt{2\pi}}\\
	\times
		\prod_{\nu a}\exp\left\{\frac{\beta}{2}\left(m_{\nu}^{a}\right)^{2}\right\} \left\langle \mathcal{N}(m_{\nu}^a;\bar{m},Q)\right\rangle _{c_1,...,c_L,f}
\end{multline}
where $\left\langle ...\right\rangle $ represents the average over $f$ and the remaining coefficients $c_k$ with $k=1,...,L$.
 We solve this model in the limit $\beta\to\infty$ in the replica-symmetric ansatz and for symmetric mixtures, namely    
\begin{equation}
    \bmu=\{\overbrace{\mu,...,\mu}^L,\overbrace{0, ... ,0}^{D-L}\}.
\end{equation}
From $\left\langle Z^{n}\right\rangle$ we obtain the free energy $\phi$ as a function of $\mu$ and other order parameters, for any value of $L$, $\alpha$ and $\alpha_D$. Then, we extremize it with respect to all the order parameters and we find the transition  by looking for the point at which $\mu$ jumps from zero to a finite value, meaning that the symmetric mixtures become a local minimum of the free energy (see appendix \ref{app:symmetric_mix} in SM for the explicit derivation). The lines in the $\alpha_D,\alpha$ space where this transition happens are plotted in Fig.~\ref{fig:phase_diagram} for $L=3,5,7$ (for binary neurons, mixtures are known to be unstable for even $L$ since \cite{amit1987information,amit1987statistical}).
For reference, in the vertical panel we also show the phase diagram of the standard Hopfield Model, showing that the lines of the Random Features Hopfield Model correctly converge to the corresponding critical points at zero temperature. 
The spinonal lines predict with good agreement the generalization transitions identified with the numerical experiments in Fig.~\ref{fig:main} and in Fig.~\ref{fig:manyL}a.

\paragraph*{Discussion.}
The ability of a Hopfield Network to store combinations of examples has been known since \cite{amit1987statistical,amit1987information}, but it gains a new framework in the light of the learning transition: if the features hidden in the examples become attractors, the network can combine them to produce new attractors that will recognize all the possible examples generated with the same features (regardless of whether they were used to train the model or not).
Note that one shortcoming of the learning phase is that, while the model is able to retrieve features if initialized close enough to one of them, there is no known way to find features if no information is known, since the model converges to a spin-glass state with high probability if initialized at random. 
The generalization phase circumvents this problem: the model builds local minima for all the possible mixtures features (up to a certain $L$), so that a test example is recognized as a stable configuration without any explicit information about the underlying features.


The scheme for generalization presented in this work could be explored in more advanced architectures. Of particular importance is the exponential version of the Hopfield Model \cite{demircigil2017model,ramsauer2020hopfield,lucibello2023exponential}: given its connection with the cross-attention mechanism, the learning and generalization transitions might play a role in the forward pass of transformers architectures \cite{vaswani2017attention}. Furthermore, the exponential Hopfield Model has been connected with diffusion models \cite{ambrogioni2023search}: since the temperature can be mapped to the diffusion time, studying the stability of mixtures of features at finite temperatures might connect to the results described in \cite{biroli2024dynamical}. 

Note that the critical lines of mixtures for the standard Hopfield Model that we reported in the upper panel of Fig.~\ref{fig:phase_diagram} were previously unknown in the literature except for their intercepts with the axes. This is because the standard calculation for symmetric mixtures leads to nonphysical results at low load and high temperature, as it was already noted in \cite{amit1987statistical}. Since it is outside of the scope of this work, we report the calculation in \cite{kalaj2024mixtures}.

On the technical side, the fact that the theory for the dense case $L_\mathrm{train}=D$ matches the numerical results in the sparse case $L_\mathrm{train}=O(1)$ is unexpected and further inquiry on this property is desirable. In particular, it might mean that an extension of the Gaussian Equivalence Theorem might hold in the case of sparse coefficients.

\paragraph*{Acknowledgements.}
MN acknowledges the support of PNRR MUR project PE0000013-FAIR. E.M.M. acknowledges the MUR-Prin 2022 funding Prot. 20229T9EAT, financed by the European Union (Next Generation EU).

\bibliographystyle{unsrt}
\bibliography{references}


\newpage

\appendix

\onecolumngrid

\tableofcontents  

\section{Signal-to-noise analysis}

In this section we derive eq.~\ref{eq::signaltonoise-mixture}. To do this, we first do the signal-to-noise analysis for the retrieval of a feature in the dense case, so that we understand the meaning of the various noise terms in a case for which we already know that the retrieval of the features happens at $\alpha_D>0$ \cite{negri2023hidden}. 

Then, we move to the sparse case and we argue that the retrieval of features is still possible and depends weakly on $L$. This exercise also allows us to set up the final case of the retrieval of a mixture of $L$ features, since the steps are the same except for the very last one.

\label{app:signal-to-noise}

\subsection{Retrieval of a feature with $L=O(D)$ coefficients}
In the fully dense case we study the signal-to-noise ratio of a single feature retrieval, being the local field 
\begin{equation}
    \begin{split}
        h_{ki}&=\sum_{j}J_{ij}f_{kj}\\
        &=\frac{1}{N}\sum_{j}\sum_{\mu}\xi_{\mu i}\xi_{\mu j}f_{kj}\\
        &=\frac{1}{ND}\sum_{k''}\sum_{k'}f_{k'i}\left[\sum_{\mu}c_{\mu k'}c_{\mu k''}\right]\left[\sum_{j}f_{k''j}f_{kj}\right]\\
	    &=\frac{1}{D}\sum_{k''}\sum_{k'}f_{k'i}\left(P\delta_{k'k''}+(1-\delta_{k'k''})O(P^{1/2})\right)\left(\delta_{k''k}+(1-\delta_{k''k})O(N^{-1/2})\right)
    \end{split}
\end{equation}
which is separated into the true signal (contributing to the retrieval) for the diagonal term, while the rest is noise. By making the orders of different contributions explicit, we have  
\begin{equation}
    \begin{split}
        h_{ki} = \frac{P}{D}\left[f_{ki}+O\left(\sqrt{\frac{D}{N}}\right)+O\left(\sqrt{\frac{D}{P}}\right)+O\left(\sqrt{\frac{D}{N}}\sqrt{\frac{D}{P}}\right)\right]
    \end{split}
\end{equation}
where we collected the order of the signal. This is compatible with the learning phase described in~\cite{negri2023hidden} for finite values of $\alpha=\frac{P}{N}$ and $\alpha_D=\frac{D}{N}$: in fact, taking $D=O(N)$ and $P=O(N)$ makes all the noise terms finite in the limit $N\to\infty$. Fixing the scaling of the lower order noises also guarantees that the higher orders one are finite in the thermodynamic limit.
\subsection{Retrieval of a feature with $L=O(1)$ coefficients}
For the sparse case, instead, 
\begin{equation}
\begin{split}
	h_{k''i}	&=\sum_{j=1}^{N}J_{ij}f_{k''j}\\
	&=\frac{1}{N}\sum_{j}\sum_{\mu}\left(\frac{1}{\sqrt{L}}\sum_{k}c_{\mu k}f_{ki}\right)\left(\frac{1}{\sqrt{L}}\sum_{k'}c_{\mu k'}f_{k'i}\right)f_{k''j}\\
	&=\frac{1}{L}\sum_{kk'}\left(\delta_{kk'}P\frac{L}{D}+(1-\delta_{kk'})O\left(\sqrt{P}\right)\frac{L^{2}}{D^{2}}\right)\left(\delta_{k'k''}+(1-\delta_{k'k''})O\left(\frac{1}{\sqrt{N}}\right)\right)f_{ki}.
\end{split}
\end{equation}
where now the patterns in $N$-dimension are properly scaled with $\sqrt{L}$ instead of $\sqrt{D}$. Solving for the four cases and rescaling with the signal (feature) order we have
\begin{equation}
    \begin{split}
        h_{k''i}	&=\frac{P}{D}\left[f_{k''i}+O\left(\sqrt{\frac{D}{N}}\right)+\frac{L}{D}O\left(\sqrt{\frac{D}{P}}\right)+\frac{L}{D}O\left(\sqrt{\frac{D^{2}}{PN}}\right)\right]
    \end{split}
\end{equation}
which provides additional insight that the learning transition also exists in the sparse case at finite $\alpha$ and that, as long as $L=O(1)$ and $N\to\infty$, $D=\alpha_D N$, $P=\alpha N$ the transition seems to weakly depend on $L$ (see also Fig.~\ref{fig:manyL}a), since the noise terms depending on $L$ vanish in this limit.
\subsection{Retrieval of a mixture of features with $L=O(1)$ coefficients}
We report the steps for eq.~\eqref{eq::signaltonoise-mixture} in the main text. Being the mixture of features defined as $\chi_{\nu i}=\sum_{k''}c_{\nu k''} f_{k'' i}$, we have
\begin{equation}
    \begin{split}
        h_{\nu i}	&=\sum_{j=1}^{N}J_{ij}\chi_{\mu j}\\
	&=\sum_{k''}c_{\mu k''}\sum_{j=1}^{N}J_{ij}f_{k''i}\\
	&=\sum_{k''}c_{\mu k''}\left[\frac{1}{N}\sum_{j}\sum_{\mu}\left(\frac{1}{\sqrt{L}}\sum_{k}c_{\mu k}f_{ki}\right)\left(\frac{1}{\sqrt{L}}\sum_{k'}c_{\mu k'}f_{k'i}\right)f_{k''j}\right]\\
	&=\sum_{k''}c_{\mu k''}\frac{P}{D}\left[f_{k''i}+O\left(\sqrt{\frac{D}{N}}\right)+\frac{L}{D}O\left(\sqrt{\frac{D}{P}}\right)+\frac{L}{D}O\left(\sqrt{\frac{D^{2}}{PN}}\right)\right]\\
	&=\frac{P}{D}\left[\sum_{k''}c_{\mu k''}f_{k''i}+O\left(\sqrt{\frac{LD}{N}}\right)+O\left(\sqrt{\frac{L^{3}}{DP}}\right)+O\left(\sqrt{\frac{L^{3}}{NP}}\right)\right].
    \end{split}
\end{equation}

\section{Extension of the Gaussian Equivalence Theorem for mixtures of features}
\label{app:gaussian_equivalence}

We are interested in the retrieval of a mixtures of features, hence, we consider the case of $L$ patterns condensed. The average of the delta function inside equation (\ref{eq:rfhmpartitionfunctioninit}) yields the probability distribution of the pattern magnetizations $m_{\nu}^a$ given the feature matrix.

Assuming that
\begin{subequations}
    \begin{align}
        \frac{1}{D}\sum_{k=1}^D f_{ki}^2&=1, \qquad\qquad \forall i,\\
    \frac{1}{\sqrt{D}}\sum_{k=1}^D f_{ki}f_{kj}&=O(1), \;\;\;\;\;\; \forall i\neq j,
    \end{align}
\end{subequations}
the \textbf{Gaussian Equivalence Theorem} \cite{mei2022generalization, gerace2020generalisation, goldt2022gaussian, goldt2020modeling, hu2022universality,Baldassi2022Learning}, in the case of mixtures of features, states that the probability distribution of the patterns magnetizations is a multivariate Gaussian
\begin{equation}
    P(\{m_{\nu}^a\} |c_{\nu 1},..., c_{\nu L},f_{ki},s_i^a)=\frac{1}{\sqrt{2\pi\det{\mathbf{Q}}}}\exp\bigg{(}-\frac{1}{2}\sum_{a,b}(m_{\nu}^a-\Bar{m}^a)(\mathbf{Q}^{-1})_{ab}(m_{\nu}^b-\Bar{m}^b)\bigg{)},
\end{equation}
where the difference with the single feature retrieval case is the mean magnetization $\Bar{m}^a$.

We can compute it using the Fourier representation of the delta function 
\begin{equation}
\begin{split}
\Bar{m}^a_\nu \equiv \big{\langle}m_{\nu}^a\big{\rangle}_c=\int \prod_{i=1}^N \frac{dv_i^{\nu}d\hat{v}_i^{\nu}}{2\pi}e^{i\sum_i v_i^{\nu}\hat{v}_i^{\nu}}&\bigg{[}\frac{1}{\sqrt{N}}\sum_i s_i^a\sigma \bigg{(}v_i^{\nu}+\frac{1}{\sqrt{D}}\sum_{k=1}^L c_{\nu k}f_{ki}\bigg{)}\bigg{]}\\
    &\times\prod_{k>L}\bigg{\langle}\exp{-i\frac{c_{\nu k}}{\sqrt{D}}\Big{(}\sum_i \hat{v}_i^{\nu}}f_{ki}\Big{)}\bigg{\rangle}_{c_{\nu k}},\\
\end{split}
\end{equation}
the coefficients matrix elements are distributed as i.i.d. Gaussian variable, therefore the average over the $c_{\nu k}$ corresponding to uncondensed features is
\begin{equation}
    \prod_{k>L}\bigg{\langle}\exp{-i\frac{c_{\nu k}}{\sqrt{D}}\Big{(}\sum_i \hat{v}_i^{\nu}}f_{ki}\Big{)}\bigg{\rangle}_{c_{\nu k}}=\exp{\bigg{[}-\frac{1}{2}\sum_{ij}\Big{(}\frac{1}{D}\sum_{k>L}f_{ki}f_{kj}\Big{)}\hat{v}_i^{\nu}\hat{v}_j^{\nu}\bigg{]}}
\end{equation}
so that
\begin{equation}
\begin{split}
   \Bar{m}^a_\nu &= \frac{1}{\sqrt{N}}\sum_is_i^a\int Dv\, \sigma \bigg{(}v_i^{\nu}+\frac{1}{\sqrt{D}}\sum_{k=1}^L c_{\nu k}f_{ki}\bigg{)}\simeq\\
    &\simeq \frac{k_0}{\sqrt{N}}\sum_i s_i^a+\frac{k_1}{\sqrt{ND}}\sum_{k=1}^L c_{\nu k}\Big{(}\sum_i f_{ki}s_i^a\Big{)}
\end{split}
\end{equation}
where we have used the definitions
\begin{subequations}
\begin{align}
     k_0 &\equiv \int Dz \, \sigma(z),  \\
     k_1 &\equiv \int Dz \, z\,\sigma(z)=\int Dz \,\sigma'(z)\\
    Dv &\equiv \frac{dv}{\sqrt{2\pi}}\exp\bigg{(}-\frac{1}{2} v^2\bigg{)} \; \bullet
\end{align}
\end{subequations}
Introducing the magnetization of the condensed features defined by
\begin{equation}
    \mu_k^a=\frac{1}{N}\sum_i f_{k i}s_i^a\,, \qquad k = 1, \dots, L
\end{equation}
and considering the case of odd activation function $\sigma(z)$ so that $k_0=0$, we get
\begin{equation}
    \Bar{m}^a_\nu = \frac{k_1}{\sqrt{\alpha_D}}\sum_{k=1}^L c_{\nu k}\mu_k^a.
\end{equation}
The covariance matrix can be shown to be the same as in the single feature retrieval case and therefore is given by

\begin{equation}
    Q_{ab}=\big{\langle}m_{\nu}^am_{\nu}^b\big{\rangle}_c - \big{\langle}m_{\nu}^a\big{\rangle}_c\big{\langle}m_{\nu}^b\big{\rangle}_c=k_*^2 q^{ab}+k_1^2 p^{ab}
\end{equation}

where
\begin{align}
    q^{ab}&=\frac{1}{N}\sum_{i=1}^N s_i^a s_i^b\\
    p^{ab}&=\frac{1}{D}\sum_{k>L}^D \mu_k^a \mu_k^b\\
    \mu_k^a&=\frac{1}{\sqrt{N}}\sum_{i=1}^N f_{ki}s_i^a,\;\;\;\;\;k>L
\end{align}

\section{Replica-symmetric calculation for mixtures of features}
\label{app:replica_calculation}

\subsection{Retrieval of a mixture of $L$ features}

We substitute the multivariate Gaussian distribution of the pattern magnetizations $m_{\nu}^a$ and insert them in the integral using delta functions, we get
\begin{equation}
\begin{split}
   \big{\langle}Z^n\big{\rangle}=& e^{-\frac{\beta}{2}nP}\sum_{\{s_i^a\}}\int\prod_{\nu a}\frac{\mathrm{d}m_{\nu}^a}{\sqrt{2\pi}}\prod_{ab}\frac{\mathrm{d}q^{ab}d\hat{q}^{ab}}{2\pi}\prod_{ab}\frac{\mathrm{d}p^{ab}d\hat{p}^{ab}}{2\pi}\prod_{k,a}\frac{\mathrm{d}\mu_k^{a}d\hat{\mu}_k^{a}}{2\pi}\\
   &\times \prod_\nu \bigg{\langle}\exp{\bigg{(}\frac{\beta}{2}\sum_a\Big{(}m_{\nu}^a+\Bar{m}^a_\nu\Big{)}^2\bigg{)}}\bigg{\rangle}_{c_{\nu 1},...,c_{\nu L}}\\
   &\times\bigg{\langle}\prod_{\nu}\frac{1}{\sqrt{2\pi\det{\mathbf{Q}}}}\exp{\bigg{(}-\frac{1}{2}\sum_{a,b}m_{\nu}^a(\mathbf{Q}^{-1})_{ab}m_{\nu}^b\bigg{)}}\\
   &\times \exp{\bigg{(}-\alpha N \sum_{a<b}\hat{q}^{ab}(q^{ab}-\frac{1}{N}\sum_i s_i^a s_i^b )-\alpha N \sum_{a\leq b}\hat{p}^{ab}(p^{ab}-\frac{1}{D}\sum_{k>L} \mu_k^a \mu_k^b)\bigg{)}}\\
   &\times \prod_{k>L,a}\exp{\bigg{(}i\hat{\mu}_k^a \Big{(}\mu_k^{a}-\frac{1}{\sqrt{N}}\sum_i f_{ki} s_i^a\Big{)}\bigg{)}}\prod_{1\leq k\leq L,a}\exp{\bigg{(}i\hat{\mu}_k^a \Big{(}\mu_k^{a}-\frac{1}{N}\sum_i f_{ki} s_i^a\Big{)}\bigg{)}}\bigg{\rangle}_{f_{ki}},\\
\end{split}
\end{equation}
we can now perform the average over the coefficients matrix.

\subsection{Average over coefficients matrix c}
The average over $c_{\nu 1},...,c_{\nu L}$ yields a different result with respect to the single feature retrieval; the term
\begin{equation}
    \bigg{\langle}\exp{\frac{\beta}{2}\sum_a\Big{(}m_{\nu}^a+\frac{k_1}{\sqrt{\alpha_D}}\sum_{k=1}^L c_{\nu k}\,\mu_k^a\Big{)}^2\bigg{\rangle}_{c_{\nu 1},...,c_{\nu L}}}
\end{equation}
can be evaluated noting that if the $c_{\nu k}$ are Gaussian distributed with zero mean and unitary standard deviation, the sum is still a random Gaussian variable with a variance that is the sum of the variances of the variables entering in the sum, so we can write
\begin{equation}
    \sum_{k=1}^L c_{\nu k}\mu_k^a=\tilde{c}_{\nu}\sqrt{\sum_{k=1}^L \big{(}\mu_k^a\big{)}^2}=\tilde{c}_{\nu} \lVert\boldsymbol{\mu}^a \rVert \quad \text{with } \tilde{c}_{\nu}\sim \mathcal{N}\big{(}0, \,1\big{)}
\end{equation}
Using the random variable $\tilde{c}_{\nu}\sim \mathcal{N}(0,1)$ we can rewrite the average as 
\begin{equation}
\begin{split}
    \bigg{\langle}\exp{\frac{\beta}{2}\sum_a\Big{(}m_{\nu}^a+\frac{k_1\,\tilde{c}_{\nu}\,}{\sqrt{\alpha_D}}\lVert\boldsymbol{\mu}^a\rVert\Big{)}^2\bigg{\rangle}_{\tilde{c}_{\nu }}} &=
    e^{\frac{\beta}{2}\sum_a(m_{\nu}^a)^2}
    \bigg{\langle}\exp{\bigg{[}\tilde{c}_{\nu }\Big{(}\frac{\beta\,k_1}{\sqrt{\alpha_D}}\sum_a m_{\nu}^a\lVert\boldsymbol{\mu}^a\rVert\Big{)}+\frac{1}{2}\tilde{c}_{\nu }^2\Big{(}\frac{\beta\,k_1^2}{\alpha_D}\sum_a\lVert\boldsymbol{\mu}^a\rVert^2\Big{)}\bigg{]}}\bigg{\rangle}_{\tilde{c}_{\nu }}\\
    &=\frac{1}{\sqrt{1-\beta\frac{k_1^2}{\alpha_D}\sum_a\lVert\boldsymbol{\mu}^a\rVert^2}}\exp{\bigg{[}\frac{\beta}{2}\sum_a(m_{\nu}^a)^2+\frac{\beta^2k_1^2}{2C\,\alpha_D}\Big{(}\sum_a m_\nu^a\lVert\boldsymbol{\mu}^a\rVert\Big{)}^2\bigg{]}}
\end{split}    
\end{equation}
where
\begin{equation}
    C=1-\beta\frac{k_1^2}{\alpha_D}\sum_a \lVert\boldsymbol{\mu}^a\rVert^2.
\end{equation}

\subsection{Integrating pattern magnetizations}

We can perform the integral over the pattern magnetizations, for a given pattern $\nu$ we have
\begin{equation}
\begin{split}
    &\frac{1}{\sqrt{\det \boldsymbol{Q}}}\int\prod_{a}\frac{\mathrm{d}m_{\nu}^a}{\sqrt{2\pi}}\exp\bigg{[}\frac{\beta}{2}\sum_a(m_{\nu}^a)^2
    +\frac{\beta^2k_1^2}{2C\,\alpha_D}\Big{(}\sum_a m_\nu^a\lVert\boldsymbol{\mu}^a\rVert\Big{)}^2
    -\frac{1}{2}\sum_{a,b}m_{\nu}^a(\mathbf{Q}^{-1})_{ab}m_{\nu}^b\bigg{]}
    \\
    &= \frac{1}{\sqrt{\det \boldsymbol{Q}}}\int\prod_{a}\frac{\mathrm{d}m_{\nu}^a}{\sqrt{2\pi}}
    \exp\bigg{[}-\frac{1}{2}\sum_{a,b} m_{\nu}^a \Big{(}-\beta\delta_{ab}-\frac{\beta^2k_1^2}{C\,\alpha_D}\lVert\boldsymbol{\mu}^a\rVert\lVert\boldsymbol{\mu}^b\rVert|+(\mathbf{Q}^{-1})_{ab}\Big{)}m_{\nu}^b\bigg{]}\\
    &= \frac{1}{\sqrt{\det \Big{(}\mathbb{I}-\beta \mathbf{Q}-\frac{\beta^2k_1^2}{C\,\alpha_D}\mathbf{p}_L \mathbf{Q}\Big{)}}}
\end{split}
\end{equation}
where we have defined
\begin{equation}
p_L^{ab}=\lVert\boldsymbol{\mu}^a\rVert \lVert\boldsymbol{\mu}^b\rVert\,.
\end{equation}
The contribution to the partition function can be written as $Z_0=e^{-\frac{\alpha N}{2}\phi_0}$, where
\begin{equation}
    \phi_0=\ln{\det{\Big{(}\mathbb{I}-\beta \mathbf{Q}-\frac{\beta^2k_1^2}{C\,\alpha_D}\mathbf{p}_L \mathbf{Q}\Big{)}}}.
    \label{eq:phi0}
\end{equation}

\subsection{Integrating the feature magnetizations}

We consider now the contribution to the partition function given by the integral over uncondensed feature magnetisation
\begin{equation}
    \begin{split}
Z_1=\int\prod_{k>L,a}\frac{d\mu_k^{a}d\hat{\mu}_k^{a}}{2\pi}&\exp{\bigg{[}\frac{\alpha}{\alpha_D}\frac{1}{2}\sum_{a,b}\hat{p}^{ab}\sum_{k>L}\mu_k^a \mu_k^b+i\sum_{k>L,a}\hat{\mu}_k^a\mu_k^a\bigg{]}}\\
    &\times \bigg{\langle}\exp{\bigg{[}-\frac{i}{\sqrt{N}}\sum_{k>L,a}\hat{\mu}_k^a\sum_i f_{ki}s_i^a\bigg{]}}\bigg{\rangle}_{f_{ki}, \,\,\,k>L},
    \end{split}
\end{equation}
for the last term we have
\begin{equation}
    \begin{split}
    \bigg{\langle}\exp{\bigg{[}-\frac{i}{\sqrt{N}}\sum_{k>L,a}\hat{\mu}_k^a\sum_i f_{ki}s_i^a\bigg{]}}\bigg{\rangle}_{f_{ki}, \,\,k>L}&=\prod_{i, k>L}\int Df_{ki}\exp{\bigg{[}-\frac{i}{\sqrt{N}}f_{ki}\sum_a \hat{\mu}_k^a s_i^a\bigg{]}} \\
    &= \exp{\bigg{[}-\frac{1}{2}\sum_{a,b}\Big{(}\sum_{k>L}\hat{\mu}_k^a \hat{\mu}_k^b\Big{)}\Big{(}\frac{1}{N}\sum_i s_i^a s_i^b\Big{)} \bigg{]}}.    
    \end{split}
\end{equation}
We can write the total contribution of the integration over feature magnetizations as $Z_1=e^{(D-L)\phi_1}$ noting that 
\begin{equation}
    \begin{split}    
    \phi_1&=\ln\int\prod_a \frac{d\hat{\mu}^a d\mu^a}{2\pi}\exp\bigg{[}i\sum_a\hat{\mu}^a \mu^a -\frac{1}{2}\sum_{a,b}\hat{\mu}^a \hat{\mu}^b q^{ab} + \frac{\alpha_T}{2}\sum_{a,b}\hat{p}^{ab}\mu^a \mu^b\bigg{]}=\\
    &=\ln\int\prod_a \frac{d\hat{\mu}^a }{\sqrt{2\pi}}\exp\bigg{[}-\frac{1}{2}\sum_{a,b}q^{ab}\hat{\mu}^a \hat{\mu}^b+\frac{1}{2\alpha_T}\sum_{a,b}\hat{\mu}^a(\mathbf{\hat{p}}^{-1})^{ab}\hat{\mu}^b+\frac{1}{2}\ln\det(\alpha_T \mathbf{\hat{p}})\bigg{]}=\\
    &=-\frac{1}{2}\ln\det(-\alpha_T \mathbf{\hat{p}})-\frac{1}{2}\ln\det(\mathbf{q}-\frac{\,\,\,\,\,\mathbf{\hat{p}}^{-1}}{\alpha_T})=\\
    &=-\frac{1}{2}\ln\det\Big{(}\mathbb{I}-\alpha_T\mathbf{q}\mathbf{\hat{p}}\Big{)},
    \label{eq:phi1}
    \end{split}
\end{equation}
the partition function now reads
\begin{equation}
    \begin{split} 
    \big{\langle}Z^n\big{\rangle} &= e^{-\frac{\beta}{2}nP}\int\prod_{ab}\frac{dq^{ab}d\hat{q}^{ab}}{2\pi}\prod_{ab}\frac{dp^{ab}d\hat{p}^{ab}}{2\pi}\prod_{a=1}^n\prod_{k=1}^L\frac{d\mu_k^{a}d\hat{\mu}_k^{a}}{2\pi}\\
    &\times \exp{\bigg{[}-\frac{\alpha N}{2}\sum_{a\neq b}\hat{q}^{ab}q^{ab}-\alpha N\sum_{a\leq b}\hat{p}^{ab}p^{ab}+\sum_{a=1}^n\sum_{k=1}^L i\hat{\mu}_k^a \mu_k^{a}\bigg{]}}\\
    &\times \exp{\bigg{[}-\frac{\alpha N}{2}\ln{\big{(}1-\beta\frac{k_1^2}{\alpha_D}\sum_a \lVert\boldsymbol{\mu}^a\rVert^2\big{)}}-\frac{\alpha N}{2}\ln{\det{\Big{(}\mathbb{I}-\beta \mathbf{Q}-\frac{\beta^2k_1^2}{C\,\alpha_D}\mathbf{p}_L \mathbf{Q}\Big{)}}}\bigg{]}}\\
    &\times \exp{\bigg{[}-\frac{\alpha_D N}{2}\ln{\det{\Big{(}\mathbb{I}-\alpha_T \mathbf{q}\mathbf{\hat{p}}}\Big{)}}\bigg{]}}\\
    &\times \bigg{\langle}\sum_{\{ s_i^a\}}\exp{\bigg{[}\frac{\alpha}{2}\sum_{a\neq b}\hat{q}^{ab}\bigg{(}\sum_i s_i^a s_i^b\bigg{)}-\frac{i}{N}\sum_{1\leq k\leq L, a}\hat{\mu}_k^a\bigg{(}\sum_i f_{k i}\,s_i^a\bigg{)}\bigg{]}}\bigg{\rangle}_{f_{ki}, 1\leq k\leq L}.
    \end{split}
\end{equation}
We do the substitution $\hat{\mu}_k^a \to iN\hat{\mu}_k^a$, so that the terms have the right scaling with $N$. 
The partition function becomes
\begin{equation}
    \begin{split} 
    \big{\langle}Z^n\big{\rangle} &= e^{-\frac{\beta}{2}nP}\int\prod_{ab}\frac{dq^{ab}d\hat{q}^{ab}}{2\pi}\prod_{ab}\frac{dp^{ab}d\hat{p}^{ab}}{2\pi}\prod_{a=1}^n\prod_{k=1}^L\frac{d\mu_k^{a}d\hat{\mu}_k^{a}}{2\pi}\\
    &\times \exp{\bigg{[}-\frac{\alpha N}{2}\sum_{a\neq b}\hat{q}^{ab}q^{ab}-\alpha N\sum_{a\leq b}\hat{p}^{ab}p^{ab}-N\sum_{a=1}^n\sum_{k=1}^L\hat{\mu}_k^a \mu_k^{a}\bigg{]}}\\
    &\times \exp{\bigg{[}-\frac{\alpha N}{2}\ln{ \left( 1-\beta\frac{k_1^2}{\alpha_D}\sum_a \lVert\boldsymbol{\mu}^a\rVert^2 \right)} -\frac{\alpha N}{2}\ln{\det{\Big{(}\mathbb{I}-\beta \mathbf{Q}-\frac{\beta^2k_1^2}{C\,\alpha_D}\mathbf{p}_L \mathbf{Q}\Big{)}}}\bigg{]}}\\
    &\times \exp{\bigg{[}-\frac{\alpha_D N}{2}\ln{\det{\Big{(}\mathbb{I}-\alpha_T \mathbf{q}\mathbf{\hat{p}}}\Big{)}}\bigg{]}}\\
    &\times \bigg{\langle}\sum_{\{ s_i^a\}}\exp{\bigg{[}\frac{\alpha}{2}\sum_{a\neq b}\hat{q}^{ab}\bigg{(}\sum_i s_i^a s_i^b\bigg{)}+\sum_{a=1}^n\sum_{k=1}^L\hat{\mu}_k^a\bigg{(}\sum_i f_{ki}\,s_i^a\bigg{)}\bigg{]}}\bigg{\rangle}_{f_{ki}, 1\leq k\leq L}.
    \end{split}
\end{equation}
We can now rewrite the last term contribution as
\begin{equation}
    Z_2=\prod_{i=1}^N\bigg{\langle}\sum_{\{ s_i^a\}}\exp{\bigg{[}\frac{\alpha}{2}\sum_{a\neq b}s_i^a\hat{q}^{ab} s_i^b+\sum_{a=1}^n s_i^a\sum_{k=1}^L\hat{\mu}_k^a \,f_{ki}\bigg{]}}\bigg{\rangle}_{f_{ki}, 1\leq k\leq L},
\end{equation}
hence, we have $Z_2 = e^{N\phi_2}$ with
\begin{equation}
    \phi_2 = \ln{\bigg{\langle}\sum_{\{ s^a\}}\exp{\bigg{[}\frac{\alpha}{2}\sum_{a\neq b}s^a\hat{q}^{ab} s^b+\sum_{a=1}^n s^a\sum_{k=1}^L\hat{\mu}_k^a \,f_{k}\bigg{]}}\bigg{\rangle}_{f_k, 1\leq k\leq L}}.
    \label{eq:phi2}
\end{equation}

\subsection{Summary expression of the partition function}
Summarizing, we defined the following quantities
\begin{equation}
    \begin{split}
        &\phi\big{(}\{\mu_k^a\}\big{)} = -\ln\big{(}1-\beta\frac{k_1^2}{\alpha_D}\sum_{a=1}^n \lVert\boldsymbol{\mu}^a\rVert^2\big{)};\\
        &\phi_0\big{(}\{\mu_k^a\},\,\mathbf{q},\,\mathbf{p} \big{)}=\ln{\det{\Big{(}\mathbb{I}-\beta \mathbf{Q}-\frac{\beta^2k_1^2}{C\,\alpha_D}\mathbf{p}_L \mathbf{Q}\Big{)}}};\\
        &\phi_1\big{(}\mathbf{q},\,\mathbf{\hat{p}} \big{)} = \ln\det\Big{(}\mathbb{I}-\alpha_T \mathbf{q}\mathbf{\hat{p}}\Big{)};\\
        &\phi_2\big{(}\{\hat{\mu}_k^a\},\,\mathbf{\hat{q}}\big{)} = \ln\bigg{\langle}\sum_{\{ s^a\}}\exp\bigg{[}\frac{\alpha}{2}\sum_{a\neq b}s^a\hat{q}^{ab} s^b+\sum_{a=1}^n s^a\sum_{k=1}^L\hat{\mu}_k^a \,f_{k}\bigg{]}\bigg{\rangle}_{f_{k}, 1\leq k\leq L};
    \end{split}
\end{equation}
where 
\begin{equation}
    \begin{split}
        \lVert\boldsymbol{\mu}^a\rVert&=\sqrt{\sum_{k=1}^L \big{(}\,\mu_k^a\,\big{)}^2};\\
        C &= 1-\beta \frac{k_1^2}{\alpha_D}\sum_{a=1}^n\lVert\boldsymbol{\mu}^a\rVert^2;\\
    p_L^{ab}&=\lVert\boldsymbol{\mu}^a\rVert\lVert\boldsymbol{\mu}^b\rVert| .
    \end{split}
    \label{eq:helpformula}
\end{equation}
Using these, we can write the complete partition function as
\begin{equation}
\begin{split}
\big{\langle}Z^n\big{\rangle}=& \int\prod_{ab}\frac{dq^{ab}d\hat{q}^{ab}}{2\pi}\prod_{ab}\frac{dp^{ab}d\hat{p}^{ab}}{2\pi}\prod_{a=1}^n\prod_{k=1}^L\frac{d\mu_k^{a}d\hat{\mu}_k^{a}}{2\pi}\\
&\times \exp{N\bigg{[}-n\frac{\alpha\beta}{2}-\frac{\alpha }{2}\sum_{a\neq b}\hat{q}^{ab}q^{ab}-\alpha \sum_{a\leq b}\hat{p}^{ab}p^{ab}-\sum_{a=1}^n\sum_{k=1}^L\hat{\mu}_k^a \mu_k^{a}\bigg{]}}\\
&\times \exp{N\bigg{[}\frac{\alpha}{2}\phi\big{(}\{\mu_k^a\}\big{)}-\frac{\alpha}{2}\phi_0\big{(}\{\mu_k^a\},\,\mathbf{q},\,\mathbf{p} \big{)}-\frac{\alpha_D}{2}\phi_1\big{(}\mathbf{q},\,\mathbf{\hat{p}} \big{)}+\phi_2\big{(}\{\hat{\mu}_k^a\},\,\mathbf{\hat{q}}\big{)}\bigg{]}}.
\label{eq:Znlast}
\end{split}
\end{equation}

\subsection{RS ansatz}

We now use the replica symmetric (RS) ansatz
\begin{equation}
    \begin{split}
       \mu_k^a&=\mu_k;\\
        \hat{\mu}_k^a&=\hat{\mu}_k;\\
        q^{ab}&=\delta^{ab}+q\,(1-\delta^{ab});\\
        \hat{q}^{ab}&=\delta^{ab}+\hat{q}\,(1-\delta^{ab});\\
        p^{ab}&=p_d\,\delta^{ab}+p\,(1-\delta^{ab});\\
        \hat{p}^{ab}&=\hat{p}_d\,\delta^{ab}+\hat{p}\,(1-\delta^{ab});\\
    \end{split}
\end{equation}
the quantities defined in equation (\ref{eq:helpformula}) become 
\begin{equation}
    \centering
    \begin{split}
    \lVert\boldsymbol{\mu}^a\rVert&=\sqrt{\sum_{k=1}^L \mu_k^2};\\
        C &= 1-\beta \frac{k_1^2}{\alpha_D}n\lVert \boldsymbol{\mu} \rVert^2\approx 1;\\
    p_L^{ab}&=\lVert \boldsymbol{\mu} \rVert^2 = \sum_{k=1}^L \mu_k^2.
    \end{split}
    \label{eq:helpformulaRS}
\end{equation}
We define for convenience the RS matrix
\begin{equation}
        Q^{ab}=Q_d\,\delta^{ab}+Q\,(1-\delta^{ab}),
\end{equation}
with the definitions
\begin{subequations}
    \begin{align}
        Q_d &= k_*^2+k_1^2\,p_d \,,\\
        Q &= k_*^2\,q+k_1^2\,p.
    \end{align}
\end{subequations}
We can evaluate the terms inside the partition function using the RS ansatz.
The term $\phi_2$ becomes
\begin{equation}
\begin{split}
\phi_2 =& \ln\bigg{\langle}e^{-\frac{\alpha}{2}n\hat{q}}\sum_{\{ s^a\}}\exp\bigg{[}\frac{\alpha}{2}\hat{q}\bigg{(}\sum_{a=1}^n s^a\bigg{)}^2
+\bigg{(}\sum_{k=1}^L\hat{\mu}_k f_k\bigg{)}\sum_{a=1}^n s^a\bigg{]}\bigg{\rangle}_{f_{k}, 1\leq k\leq L}=\\
=&-\frac{\alpha n}{2}\hat{q}+\ln\bigg{\langle}\sum_{\{ s^a\}}\int \mathrm{D}z\exp \bigg{[}\bigg{(}\sqrt{\alpha \hat{q}}z+\sum_{k=1}^L\hat{\mu}_k f_k\bigg{)}\sum_{a=1}^n s^a\bigg{]}\bigg{\rangle}_{f_{k}, 1\leq k\leq L}=\\
=&-\frac{\alpha n}{2}\hat{q}+n\bigg{\langle}\int \mathrm{D}z \ln 2\cosh \bigg{(}\sqrt{\alpha \hat{q}}z+\sum_{k=1}^L\hat{\mu}_k f_k\bigg{)}\bigg{\rangle}_{f_{k}, 1\leq k\leq L}.  
\end{split}
\end{equation}
We can rewrite the quantity (\ref{eq:phi0}) as $\phi_0=\ln\det \mathbf{D}_0$ with 
\begin{equation}
\begin{split}
D_0^{ab}=&\delta^{ab}-\beta Q^{ab}-\frac{\beta^2 k_1^2}{C\alpha_D}\big{(}\,\mathbf{p}_L \,\mathbf{Q}\,\big{)}^{ab}\\
=&\delta^{ab}-\beta\bigg{(}Q_d \delta^{ab}+Q\big{(}1-\delta^{ab}\big{)}\bigg{)}-\frac{\beta^2 k_1^2}{C\alpha_D}\bigg{(}\sum_{k=1}^L \mu_k^2\bigg{)}\sum_{c=1}^n\bigg{(}Q_d\delta^{cb}+Q\big{(}1-\delta^{cb}\big{)}\bigg{)}=\\
=&\delta^{ab}\bigg{[}1-\beta\big{(}Q_d-Q\big{)}\bigg{]}-\beta Q-\frac{\beta^2k_1^2}{C\alpha_D}\bigg{[}Q_d+\big{(}n-1\big{)}Q\bigg{]}\bigg{(}\sum_{k=1}^L\mu_k^2\bigg{)},
\label{eq:D0}
\end{split}
\end{equation}
where in the last line we use the RS ansatz for the features magnetizations.

For a generic $n\times n$ replica symmetric matrix of the form 
\begin{equation}
    X^{ab}=x_d \delta^{ab}+x\big{(}1-\delta^{ab}\big{)},
    \label{eq:RSmatrix}
\end{equation}
holds the formula
\begin{equation}
    \ln\det \mathbf{X}=n\ln\Big{(}x_d-x\Big{)}+n\frac{x}{x_d-x}+O\big{(}n^2\big{)},
    \label{eq:lndetRSmatrix}
\end{equation}
using it in equation (\ref{eq:D0}) and considering that in the limit $n\to0$ we have $C\approx 1$ we get
\begin{equation}
    \phi_0=n\ln\Big{[}1-\beta\big{(}Q_d-Q\big{)}\Big{]}-n\frac{\beta Q+\frac{\beta^2 k_1^2}{\alpha_D}\big{(}Q_d-Q\big{)}\bigg{(}\sum_{k=1}^L\mu_k^2\bigg{)}}{1-\beta\big{(}Q_d-Q\big{)}}.
\end{equation}

We can write in a similar way $\phi_1=-\frac{1}{2}\ln\det \mathbf{D}_1$, with
\begin{equation}
    \mathbf{D}_1=\mathbb{I}-\alpha_T \,\mathbf{q}\mathbf{\hat{p}},
    \label{eq:D1}
\end{equation}
using the RS ansatz we get
\begin{equation}
\begin{split}
    D_1^{ab}=&\delta^{ab}-\alpha_T \sum_{c=1}^n q^{ac}\hat{p}^{cb}=\\
    =&\delta^{ab}-\alpha_T\sum_{c=1}^n\Big{(}\delta^{ac}+q\big{(}1-\delta^{ac}\big{)}\Big{)}\Big{(}\hat{p}_d\delta^{cb}+\hat{p}\big{(}1-\delta^{cb}\big{)}\Big{)}=\\
    =&\delta^{ab}-\alpha_T\sum_{c=1}^n\Big{(}\big{(}1-q\big{)}\delta^{ac}+q\Big{)}\Big{(}\big{(}\hat{p}_d-\hat{p}\big{)}\delta^{cb}+\hat{p}\Big{)}=\\
    =&\delta^{ab}-\alpha_T\bigg{[}\big{(}1-q\big{)}\big{(}\hat{p}_d-\hat{p}\big{)}\delta^{ab}+\big{(}1-q\big{)}\hat{p}+q\big{(}\hat{p}_d-\hat{p}\big{)}+nq\hat{p}\bigg{]}=\\
    =&\delta^{ab}-\alpha_T\bigg{[}\delta^{ab}\Big{(}\hat{p}_d-\hat{p}-q\hat{p}_d+q\hat{p}\Big{)}+\hat{p}+q\hat{p}_d+\Big{(}n-2\Big{)}q\hat{p}\bigg{]}=\\
    =&\delta^{ab}\bigg{(}1-\alpha_T\Big{(}\hat{p}_d + \big{(}n-1\big{)}q\hat{p}\Big{)}\bigg{)}-\alpha_T\Big{(}1-\delta^{ab}\Big{)}\bigg{(}\hat{p}+q\hat{p}_d+\big{(}n-2\big{)}q\hat{p}\bigg{)},
\end{split}
\end{equation}
using formula (\ref{eq:lndetRSmatrix}) we get
\begin{equation}
    \phi_1=n\ln \bigg{(}1-\alpha_T\big{(}\hat{p}_d-\hat{p}\big{)}\big{(}1-q\big{)}\bigg{)}-n\alpha_T\frac{\hat{p}+q\hat{p}_d-2q\hat{p}}{1-\alpha_T\big{(}\hat{p}_d-\hat{p}\big{)}\big{(}1-q\big{)}}.
\end{equation}
The term $\phi$ in the RS ansatz becomes 
\begin{equation}
\begin{split}
    \phi =& -\ln\big{(}1-\beta\frac{k_1^2}{\alpha_D}\sum_{a=1}^n \lVert\boldsymbol{\mu}^a\rVert^2\big{)}=\\
    =&-\ln\big{(}1-\beta\frac{k_1^2}{\alpha_D} n \sum_{k=1}^L\mu_k^2\big{)}=\\
    =& n\beta\frac{k_1^2}{\alpha_D}\sum_{k=1}^L \mu_k^2 + O\big{(}n^2\big{)}.
\end{split}
\end{equation}
Finally, the remaining terms in eq.~\ref{eq:Znlast} become
\begin{align}
    \sum_{a\neq b}q^{ab}\hat{q}^{ab}=&n\big{(}n-1\big{)}q\hat{q}=-nq\hat{q}+O\big{(}n^2\big{)};\\
\sum_{a,b}p^{ab}\hat{p}^{ab}=&n\big{(}\hat{p}_d p_d-\hat{p}p\big{)};\\
    \sum_{a=1}^n\sum_{k=1}^L \hat{\mu}_k^a\mu_k^a=&n\sum_{k=1}^L\hat{\mu}_k\mu_k.
\end{align}
The partition function (\ref{eq:Znlast}) now reads
\begin{equation}
    \big{\langle}Z^n\big{\rangle}=\int \frac{dqd\hat{q}}{2\pi}\frac{dpd\hat{p}}{2\pi}dp_d d\hat{p}_d\prod_{k=1}^L\frac{d\mu_k d\hat{\mu}_k}{2\pi}\\exp\Big{(} -\beta N f\big{(}q,\hat{q}, p, \hat{p},p_d, \hat{p}_d,\{\mu_k\},\{\hat{\mu}_k\}\big{)}\Big{)},
\end{equation}
where the function at the exponent is defined as
\begin{equation}
    -\beta f\big{(}q,\hat{q}, p, \hat{p},p_d, \hat{p}_d,\{\mu_k\},\{\hat{\mu}_k\}\big{)} = \lim_{\substack{n\to0\\N\to+\infty}}\frac{1}{N}\frac{\big{\langle}Z^n\big{\rangle}-1}{n};
\end{equation}
we have
\begin{equation}
\begin{split}
    &-\beta f\big{(}q,\hat{q}, p, \hat{p},p_d, \hat{p}_d,\{\mu_k\},\{\hat{\mu}_k\}\big{)} =\\
    =& -\beta\frac{\alpha}{2}-\frac{\alpha}{2}\hat{q}\big{(}1-q\big{)}-\frac{\alpha}{2}\hat{p}_d p_d+\frac{\alpha}{2}p\hat{p}-\sum_{k=1}^L\hat{\mu}_k\mu_k+\alpha\beta\frac{k_1^2}{2\alpha_D}\sum_{k=1}^L\mu_k^2+\\
    &-\frac{\alpha}{2}\bigg{[}\ln\Big{[}1-\beta\big{(}Q_d-Q\big{)}\Big{]}-\frac{\beta Q+\frac{\beta^2 k_1^2}{\alpha_D}\big{(}Q_d-Q\big{)}\big{(}\sum_{k=1}^L\mu_k^2\big{)}}{1-\beta\big{(}Q_d-Q\big{)}}\bigg{]}+\\
    &-\frac{\alpha_D}{2}\bigg{[} \ln \bigg{(}1-\alpha_T\big{(}\hat{p}_d-\hat{p}\big{)}\big{(}1-q\big{)}\bigg{)}-\alpha_T\frac{\hat{p}+q\hat{p}_d-2q\hat{p}}{1-\alpha_T\big{(}\hat{p}_d-\hat{p}\big{)}\big{(}1-q\big{)}}\bigg{]}+\\
    &+\bigg{\langle}\int \mathrm{D}z \ln 2\cosh \bigg{(}\sqrt{\alpha \hat{q}}z+\sum_{k=1}^L\hat{\mu}_k f_k\bigg{)}\bigg{\rangle}_{f_{k}, 1\leq k\leq L}
\end{split}    
\end{equation}
We impose the following scaling with the inverse temperature $\beta$:
\begin{subequations}
\begin{align}
    \hat{q}&\to\beta^2\hat{q};\\
    \hat{\mu}_k&\to\beta\hat{\mu}_k;\\
    \hat{p}&\to\beta\hat{p};\\
    \hat{p}_d&\to\beta\hat{p}_d;
\end{align}
\end{subequations}
the resulting RS free energy is
\begin{equation}
    \begin{split}
    f^{RS}=&+\frac{\alpha}{2}+\frac{\alpha}{2}\hat{q}\beta\big{(}1-q\big{)}+\frac{\alpha}{2}\hat{p}_d p_d-\frac{\alpha}{2}p\hat{p}+\sum_{k=1}^L\hat{\mu}_k\mu_k-\frac{k_1^2}{2}\frac{\alpha}{\alpha_D}\sum_{k=1}^L\mu_k^2\\
    &+\frac{\alpha}{2\beta}\bigg{[}\ln\Big{[}1-\beta\big{(}Q_d-Q\big{)}\Big{]}-\frac{\beta Q+\frac{\beta^2 k_1^2}{\alpha_D}\big{(}\sum_{k=1}^L\mu_k^2\big{)}\big{(}Q_d-Q\big{)}}{1-\beta\big{(}Q_d-Q\big{)}}\bigg{]}\\
    &+\frac{\alpha_D}{2\beta}\bigg{[} \ln \bigg{(}1-\alpha_T\big{(}\hat{p}_d-\hat{p}\big{)}\beta\big{(}1-q\big{)}\bigg{)}-\alpha_T\beta\frac{\hat{p}+q\hat{p}_d-2q\hat{p}}{1-\alpha_T\big{(}\hat{p}_d-\hat{p}\big{)}\beta\big{(}1-q\big{)}}\bigg{]}\\
    &-\frac{1}{\beta}\bigg{\langle}\int Dz \ln 2\cosh\beta \bigg{(}\sqrt{\alpha \hat{q}}z+\sum_{k=1}^L\hat{\mu}_k f_k\bigg{)}\bigg{\rangle}_{f_{k}, 1\leq k\leq L},   
    \label{eq:rfhmf_rs}
    \end{split}
\end{equation}
where the order parameters are the solutions of the saddle-point equations.

\subsection{Saddle-point equations}

We can now search for the saddle-point equations deriving the free energy with respect to the order parameters, the final result is:
\begin{subequations}
\begin{align}
\mu_k &= \left\langle f_k\tanh\beta \bigg{(}\sqrt{\alpha \hat{q}}z+\sum_{k=1}^L\hat{\mu}_k f_k \bigg{)} \right\rangle_{f,z};\\
    q &= \left\langle\tanh^2\beta \bigg{(}\sqrt{\alpha \hat{q}}z+\sum_{k=1}^L\hat{\mu}_k f_k\bigg{)} \right\rangle_{f,z};\\
    p&=\frac{1}{\alpha_D}\sum_{k=1}^L\mu_k^2+\frac{q+\frac{\alpha}{\alpha_D}\hat{p}\beta\big{(}1-q\big{)}^2}{\Big{(}1-\frac{\alpha}{\alpha_D}\beta\big{(}1-q\big{)}\big{(}\hat{p}_d-\hat{p}\big{)}\Big{)}^2};\\
    p_d&=\frac{1}{\alpha_D}\sum_{k=1}^L\mu_k^2+\frac{1+\frac{\alpha}{\alpha_D}\beta\big{(}2\hat{p}-\hat{p}_d\big{)}\big{(}1-q\big{)}^2}{\Big{(}1-\frac{\alpha}{\alpha_D}\beta\big{(}1-q\big{)}\big{(}\hat{p}_d-\hat{p}\big{)}\Big{)}^2};\\
    \hat{q}&=\frac{k_*^2\big{(}k_1^2 p+k_*^2 q\big{)}}{\Big{(}1-\beta k_1^2\big{(}p_d-p\big{)}-\beta k_*^2\big{(}1-q\big{)}\Big{)}^2}+
    \frac{\hat{p}+\frac{\alpha}{\alpha_D}\beta q\big{(}\hat{p}_d-\hat{p}\big{)}^2}{\beta\Big{(}\frac{\alpha}{\alpha_D}\beta\big{(}1-q\big{)}\big{(}\hat{p}_d-\hat{p}\big{)}\Big{)}^2};\\
    \hat{p}&=\frac{\beta k_1^2\big{(}k_1^2 p+k_*^2 q\big{)}}{\Big{(}1-\beta k_1^2\big{(}p_d-p\big{)}-\beta k_*^2\big{(}1-q\big{)}\Big{)}^2};\\
    \hat{p}_d&=\frac{k_1^2\big{(}1+\beta k_1^2\big{(}2p-p_d\big{)}+\beta k_*^2 \big{(}2q-1\big{)}\big{)}}{\Big{(}1-\beta k_1^2\big{(}p_d-p\big{)}-\beta k_*^2\big{(}1-q\big{)}\Big{)}^2};\\
    \hat{\mu}_k&=\frac{\alpha}{\alpha_D}\Big{(}\hat{p}_d-\hat{p}\Big{)}\mu_k.
    \label{eq:rfhmspeq}
\end{align}
\end{subequations}

\section{Specialization for symmetric mixtures}
\label{app:symmetric_mix}
We focus on symmetric mixtures of features defined by
\begin{equation}
\boldsymbol{\mu} = \mu_L\big{(} \overbrace{1,\, \dotsc, 1}^L,\,0,\, \dotsc, 0\,\big{)}.
\label{eq:feamix}
\end{equation}
Note that, in the main text, for clarity we used $\mu$ instead of $\mu_L$ in the definition of symmetric mixtures. Here we keep the subscript $L$ to stress that the solution depends on the number of features in the mixture.

Substituting this ansatz into the saddle point equations (\ref{eq:rfhmspeq}) we get
\begin{subequations}
\begin{align}
    \mu_L &= \frac{1}{L} \left\langle \zeta_L\tanh\beta\Big{(}\sqrt{\alpha \hat{q}}z+\hat{\mu}_L\zeta_L\Big{)} \right\rangle _{z,\zeta_L};\\
    q &= \left\langle\tanh^2\beta\Big{(}\sqrt{\alpha \hat{q}}z+\hat{\mu}_L\zeta_L\Big{)} \right\rangle_{z,\zeta_L};\\
    \hat{\mu}_L&=\frac{\alpha}{\alpha_D}\Big{(}\hat{p}_d-\hat{p}\Big{)}\mu_L;\\
    p&=\frac{L}{\alpha_D}\mu_L^2+\frac{q+\frac{\alpha}{\alpha_D}\beta\hat{p}\big{(}1-q\big{)}^2}{\Big{(}1-\frac{\alpha}{\alpha_D}\beta\big{(}1-q\big{)}\big{(}\hat{p}_d-\hat{p}\big{)}\Big{)}^2};\\
    p_d&=\frac{L}{\alpha_D}\mu_L^2+\frac{1+\frac{\alpha}{\alpha_D}\beta\big{(}2\hat{p}-\hat{p}_d\big{)}\big{(}1-q\big{)}^2}{\Big{(}1-\frac{\alpha}{\alpha_D}\beta\big{(}1-q\big{)}\big{(}\hat{p}_d-\hat{p}\big{)}\Big{)}^2};
\end{align}
\end{subequations}
where $\zeta_L =\sum_{k=1}^L f_k$ is the sum of feature components, which follows a binomial distribution defined as
\begin{equation}
    P(\zeta_L)=2^{-L}\binom{L}{k}, \qquad \text{with} \;\; k = (\zeta_L+L)/2.
    \label{eq:zetaprob}
\end{equation}
The other saddle point equations do not depend on the particular ansatz chosen for the feature magnetizations.

Substituting the ansatz inside the RS free energy expression (\ref{eq:rfhmf_rs}) we get for a $L$-mixture of features
\begin{equation}
    \begin{split}
f_n^{RS}=&+\frac{\alpha}{2}+\frac{\alpha}{2}\hat{q}\beta\big{(}1-q\big{)}+\frac{\alpha}{2}\hat{p}_d p_d-\frac{\alpha}{2}p\hat{p}+L\hat{\mu}_L\mu_L-L\frac{k_1^2}{2}\frac{\alpha}{\alpha_D}\mu_L^2+\nonumber\\
    &+\frac{\alpha}{2\beta}\bigg{[}\ln\Big{[}1-\beta\big{(}Q_d-Q\big{)}\Big{]}-\frac{\beta Q+L\frac{\beta^2 k_1^2}{\alpha_D}\mu_L^2\big{(}Q_d-Q\big{)}}{1-\beta\big{(}Q_d-Q\big{)}}\bigg{]}+\nonumber\\
    &+\frac{\alpha_D}{2\beta}\bigg{[} \ln \bigg{(}1-\alpha_T\big{(}\hat{p}_d-\hat{p}\big{)}\beta\big{(}1-q\big{)}\bigg{)}-\alpha_T\beta\frac{\hat{p}+q\hat{p}_d-2q\hat{p}}{1-\alpha_T\big{(}\hat{p}_d-\hat{p}\big{)}\beta\big{(}1-q\big{)}}\bigg{]}+\nonumber\\
    &-\frac{1}{\beta}\bigg{\langle}\int Dz \ln 2\cosh\beta \bigg{(}\sqrt{\alpha \hat{q}}z+\hat{\mu}_L \zeta_L\bigg{)}\bigg{\rangle}_{\zeta_L}.    
    \end{split}
\end{equation}

\subsection{Zero-temperature limit}
We will focus on the $T\to0$ limit that can be performed defining the quantities:
\begin{subequations}
\begin{align}
    \delta q&=\beta(1-q);\\
    \delta p&=\beta(p_d-p);\\
    \delta\hat{p}&=\hat{p}_d-\hat{p};\\
    \delta\hat{p}_d&=\frac{\hat{p}_d+\hat{p}}{2\beta}.
\end{align}
\end{subequations}

The resulting equations are 
\begin{subequations}
\begin{align}
    \mu_L&=\frac{1}{L}\Big{\langle}\zeta_L \erf{\Big{(}\frac{\hat{\mu}_L\zeta_L}{\sqrt{2\alpha \hat{q}}}\Big{)}}\Big{\rangle}_{\zeta_L};\\
    \delta q&=\frac{2}{\sqrt{\alpha \hat{q}}}\Big{\langle} G\Big{(}\frac{\hat{\mu}_L\zeta_L}{\sqrt{\alpha \hat{q}}}\Big{)}\Big{\rangle}_{\zeta_L};\\
    \delta p&=\frac{\delta q}{1-\frac{\alpha}{\alpha_D}\delta q\delta \hat{p}};\\
    p_d&=\frac{L}{\alpha_D}\mu_L^2+\frac{1+\frac{\alpha}{\alpha_D}\delta \hat{p}_d\delta q^2}{\big{(}1-\frac{\alpha}{\alpha_D}\delta q\delta \hat{p}\big{)}^2};\\
    \hat{\mu}_L&=\frac{\alpha}{\alpha_D}\delta \hat{p}\,\mu_L;\\
    \hat{q}&=\frac{k_*^2\big{(}k_1^2 p_d+k_*^2 \big{)}}{\big{(}1-k_1^2\delta p-k_*^2\delta q\big{)}^2}+
    \frac{\delta\hat{p}_d+\frac{\alpha}{\alpha_D}\delta\hat{p}^2}{\big{(}1-\frac{\alpha}{\alpha_D}\delta q\delta \hat{p}\big{)}^2};\\
    \delta\hat{p}&=\frac{k_1^2}{1-k_1^2\delta p-k_*^2\delta q};\\
    \delta\hat{p}_d&=\frac{k_1^2\big{(}k_1^2 p_d+k_*^2\big{)}}{\big{(}1-k_1^2\delta p-k_*^2\delta q\big{)}^2};
    \label{eq:speqT=0}
\end{align}
\end{subequations}
where we have used the functions defined by
\begin{subequations}
\begin{align}
    \erf\big{(}x\big{)}&=\frac{2}{\sqrt{\pi}}\int_0^{+\infty}\mathrm{d}t\,e^{-t^2};\\
    G\big{(}x\big{)}&=\frac{1}{\sqrt{2\pi}}e^{-\frac{x^2}{2}}.
\end{align}
\end{subequations}

\subsection{Numerical analysis of saddle-point equations}\label{sec:numericalanalysisrfhm}
The saddle-point equations (\ref{eq:speqT=0}) are solved numerically using the following iterative method. 

\subsubsection{Iterative method}
Suppose we have a system of $M$ non-linear coupled equations for the order parameters $x_i$ with external parameters $\boldsymbol{\vartheta}$, given by:
\begin{equation}
    x_i = g_i\big{(}\,\vect{x};\, \boldsymbol{\vartheta}\big{)},
    \label{eq:iter}
\end{equation}
with $i = 1,\, \dotsc, \, M$. The iterative process for fixed external parameters $\vect{\vartheta}$ unfolds as follows:
\begin{enumerate}
    \item Begin with an initial guess $\vect{x}^{(0)}$.
    \item Compute order parameters at time $t$ using the rule
        \begin{equation}
            x_i^{(t)} = \gamma x_i^{(t-1)} + (1-\gamma)\, g_i\big{(}\,\vect{x}^{(t-1)};\vect{\vartheta}\big{)},
        \label{eq:rs_iter}
        \end{equation}
        where $\gamma\in [0,1)$ serves as an hyperparameter for damping, enhancing convergence.
    \item Evaluate $\delta=|\vect{x}^{(t)}-\vect{x}^{(t-1)}|$.
    \item If $\delta < \epsilon$ stop the procedure; otherwise, restart from step $2$.
\end{enumerate}
A fixed point $\vect{x}^*$ of this method is also a solution of the set of equations (\ref{eq:iter}). The convergence of this procedure depends significantly on the initial point. Therefore, it proves helpful to know the theoretical solution for some specific values of the external parameters. Thereafter we can systematically explore the parameter space with small steps, employing the previously identified fixed point as our initial guess.

\subsubsection{Phase diagram}
With this method, we build the spinodal lines as the locus of points where the solution with $\mu>0$ becomes unstable.

In this case we span the phase diagram with horizontal cuts corresponding to a fixed value of $\alpha$ and varying the value of $\alpha_D$, starting from $\alpha_D=0$ we find the ferromagnetic solution with the iterative method using an initial guess with $\mu>0$.

Then, we increase $\alpha_D$ by a small quantity $\delta\alpha_D$ and repeat the procedure using the previously found fixed point as initial guess for the new iteration. For a certain value $\alpha_D(\alpha)$ the ferromagnetic state is no more a fixed point of the iterative method signaling the developing of instability. 

\end{document}